\documentclass[aps,pra,showpacs,superscriptaddress,preprint]{revtex4}
\usepackage{amsmath}
\usepackage{graphicx}

\catcode`ð=\active
 \defð{\u{g}}
 \catcode`Ð=\active
 \defÐ{\u{G}}
 \catcode`Ý=\active
\defÝ{\. I}
 \catcode`ö=\active
\defö{\"{o}}
 \catcode`Ö=\active
 \defÖ{\"O}
 \catcode`ü=\active
 \defü{\"{u}}
 \catcode`Ü=\active
 \defÜ{\"{U}}
 \catcode`Þ=\active
\defÞ{\c{S}}
 \catcode`þ=\active
 \defþ{\c{s}}
 \catcode`ý=\active
 \defý{{\i}}
 \catcode`ç=\active
\defç{\d{c}}
 \catcode`Ç=\active
\defÇ{\d{C}}

\begin{document}

\title{\textbf{Bound states of the Klein-Gordon equation for vector and
scalar general Hulth\'{e}n-type potentials in }$D$\textbf{-dimension }}
\author{{\small Sameer M. Ikhdair}}
\email[E-mail: ]{sikhdair@neu.edu.tr}
\affiliation{Department of Physics, Near East University, Nicosia, North Cyprus, Turkey}
\date{%
\today%
}

\begin{abstract}
We solve the Klein-Gordon equation in any $D$-dimension for the scalar and
vector general Hulth\'{e}n-type potentials with any $l$ by using an
approximation scheme for the centrifugal potential. Nikiforov-Uvarov method
is used in the calculations. We obtain the bound state energy eigenvalues
and the corresponding eigenfunctions of spin-zero particles in terms of
Jacobi polynomials. The eigenfunctions are physical and the energy
eigenvalues are in good agreement with those results obtained by other
methods for $D=1$ and $3$ dimensions. Our results are valid for $q=1$ value
when $l\neq 0$ and for any $q$ value when $l=0$ and $D=1$ or $3$. The $s$%
-wave ($l=0$) binding energies for a particle of rest mass $m_{0}=1$ are
calculated for the three lower-lying states $(n=0,1,2)$ using pure vector
and pure scalar potentials.

\end{abstract}

\maketitle

\newpage

\section{Introduction}

In nuclear and high energy physics [1,2], it is necessary to obtain the
exact bound state energy spectrum of non-relativistic or relativistic wave
equations for various potentials. In relativistic quantum mechanics, the
Klein-Gordon and Dirac wave equations are most frequently used to understand
the dynamics of a particle in large class of potentials. Therefore, much
works have been done to solve these relativistic wave equations for various
radial and angular potentials. Unfortunately, some quantum mechanical
equations with these potentials can be solved analytically only for $s$-wave
equation (i.e., $l=0$ states). Some other works have been done to study the $%
l\neq 0$ bound state solutions of the Klein-Gordon equation with pure
scalar, pure vector and mixed vector and scalar potentials in various
dimensional space using different methods. The scalar coupling constant is
almost taken to be equal to the vector coupling constant (i.e., $S_{0}=V_{0}$%
) [2-14]. For this case$,$ the Klein-Gordon equation reduces to a Schr\"{o}%
dinger-like equation and thereby the bound state solutions are easily
obtained by using the well-known methods developed in nonrelativistic
quantum mechanics [2]. On the other hand, it has been shown that when $%
S_{0}\geq V_{0}$, there exist real bound state solutions (i.e., the radial
wave function must satisfy the boundary condition that it becomes zero when $%
r\rightarrow \infty $ and also finite at $r=0)$. However, there are very few
exact solvable cases [15-19]. The bound state solutions for the last case is
obtained for the $s$-wave Klein-Gordon equation with the exponential [15]
and some other classes of potentials [16]. Chen \textit{et al.} [19] used
the exponential function transformation approach along with an approximation
for the centrifugal potential to find bound-state solutions of the
Klein-Gordon equation with the vector and scalar Hulth\'{e}n potentials for
different $l.$

The $D$-dimensional Schr\"{o}dinger and Klein-Gordon wave equations have
been solved for various types of angular and radial dependent potentials
using the Nikiforov-Uvarov (N-U) method [20-25]. The exact bound state
solutions for the Schr\"{o}dinger, Klein-Gordon and Salpeter wave equations,
in one-dimension ($1D$), with the generalized Woods-Saxon potential are also
obtained [26-29]. These solutions of the $D$-dimensional Klein-Gordon
equation describing a spin-zero particle are obtained for the case $%
V_{0}=S_{0}$ ring-shaped Kratzer-type and pseudoharmonic potentials [21,22]$%
. $ Furthermore, analytic solution of radial Schr\"{o}dinger equation has
been given for the $l\neq 0$ general Hulth\'{e}n potential within an
approximation to the centrifugal barrier term [25,29]$.$ Simsek and Egrifes
[30] have investigated the reality of exact bound states of the $1D$
Klein-Gordon equation with complex and/or PT-symmetric non-Hermitian
exponential-type like general Hulth\'{e}n potential. Berkdemir [31] has also
applied the method to solve the Klein-Gordon equation of a spin-zero
particle for $V(r)=S(r)$ Kratzer-type potentials. Qiang \textit{et al.} [32]
have given an approximate analytic solution of the $l\neq 0$ Klein-Gordon
equation for the general Hulth\'{e}n-type potential within the approximation
given in [29]. Saad [33] has presented an approximate solution of the $l\neq
0$ bound states of the Klein-Gordon equation in $D$-dimensions with the
general Hulth\'{e}n-type potentials following the model used in [29,32].
Chen \textit{et al.} [34] have employed two semiclassical methods to
determine the bound state energy spectrum of Klein-Gordon equation when $%
S_{0}\geq V_{0}$. Dong et al [35] have also obtained the bound state
solutions of the Schr\"{o}dinger equation for an exponential-type potential
following Refs. [29,32]. Very recently, we have obtained the bound-state
solutions for the $D$-dimensional Schr\"{o}dinger equation with the
Manning-Rosen potential [36] which can be reduced to the Hulth\'{e}n
potential and the exponential-type hyperbolic potential [37] using a novel
approximation to the centrifugal term [29,36,37].

The purpose of the present paper is to study the relativistic
characteristics of the scalar and vector general Hulth\'{e}n-type potentials
written, respectively, as follows [32,33]:%
\begin{equation}
V(r)=-\frac{V_{0}e^{-r/r_{0}}}{1-qe^{-r/r_{0}}},\text{ }S(r)=-\frac{%
S_{0}e^{-r/r_{0}}}{1-qe^{-r/r_{0}}},\text{ }r_{0}=\alpha ^{-1},\text{ }q\neq
0\text{ }
\end{equation}%
where $V_{0}$ and $S_{0}$ represent the coupling constants of the vector and
scalar general Hulth\'{e}n-type potentials, respectively, $\alpha $ is the
screening range parameter, $r_{0}$ represents the spatial range and $q$ is
the deformation parameter [26-28]. This potential has been widely used in a
number of areas of physics. In atomic physics, $V_{0}=Ze^{2}/r_{0}$ where $Z$
is the atomic number. Equation (1) behaves like a Coulomb potential $%
V_{C}(r)=-Ze^{2}/r$ when $r\ll r_{0}$, but decreases exponentially in the
asymptotic region when $(r\gg r_{0}),$ so its capacity for bound state is
smaller than the Coulomb potential. The Klein-Gordon equation has been
solved analytically in [30] only for the $s$-wave. This equation cannot be
solved analytically for $l\neq 0$ because of the centrifugal term, $1/r^{2}.$
Therefore, for $l\neq 0,$ we must use an approximation for the centrifugal
term similar to other authors [29,38-43] to find approximate analytical
solutions of the $D$-dimensional Klein-Gordon equation with vector and
scalar general Hulth\'{e}n-type potentials. However, the resulting analytic
solutions are only valid for $q=1$ in the $l\neq 0$ case and for any $q$
value in the $l=0$ and $D=1,3$.

This paper is organized as follows: In Section II, we will derive $l\neq 0$
state solutions within the approximation given in [29] for the $D$%
-dimensional Klein-Gordon equation describing spin-zero particle with the
vector and scalar general Hulth\'{e}n-type potentials. We further separate
the wave equation into radial and angular parts. Section III is devoted to a
brief description of the N-U method. In Section IV, we present general
solutions to the radial and angular equations in $D$-dimensions. We also
give discussions of the solution in various dimensions and the constraints
for obtaining the real bound state solutions. Further, numerical values for
the three lower-lying states when pure vector and pure scalar potentials are
used with coupling constants $V_{0}=0.25$ and $S_{0}=0.25$ and unity mass
particle. Finally, the relevant conclusions are given in Section V.

\section{The Klein-Gordon Equation with Scalar and Vector Potentials}

The time-independent Klein-Gordon equation (in any arbitrary $D$-dimension)
with scalar $S(r)$ and vector $V(r)$ potentials$,$ $r=\left\vert \mathbf{r}%
\right\vert ,$describing spin-zero particle of rest mass, $m_{0},$ can be
written as (in the relativistic natural units $\hbar =c=1$) [2,21,22]%
\begin{equation}
\left\{ \mathbf{\nabla }_{D}^{2}+\left[ E_{nl}-V(r)\right] ^{2}-\left[
m_{0}+S(r)\right] ^{2}\right\} \psi _{l_{1}\cdots l_{D-2}}^{(l_{D-1}=l)}(%
\mathbf{x})=0,\text{ }\nabla _{D}^{2}=\sum\limits_{j=1}^{D}\frac{\partial
^{2}}{\partial x_{j}^{2}},
\end{equation}%
where $E_{nl}$ denotes the relativistic energy and $\mathbf{\nabla }_{D}^{2}$
denotes the $D$-dimensional Laplacian. Usually, it is required that $%
S_{0}\geq V_{0}$ and $m_{0}>E_{nl}$ for the existence of real bound states
[15-17]. Furthermore, the $\mathbf{x}$ is a $D$-dimensional position vector,
with the unit vector along $\mathbf{x}$ is usually denoted by $\widehat{%
\mathbf{x}}=\mathbf{x}/r,$ is the hyperspherical Cartesian components $%
x_{1},x_{2},\cdots ,x_{D}$ given as follows (cf. Refs. [20-25] and the
references therein):%
\begin{equation*}
x_{1}=r\cos \theta _{1}\sin \theta _{2}\cdots \sin \theta _{D-1},
\end{equation*}%
\begin{equation*}
x_{2}=r\sin \theta _{1}\sin \theta _{2}\cdots \sin \theta _{D-1},
\end{equation*}%
\begin{equation*}
x_{3}=r\cos \theta _{2}\sin \theta _{3}\cdots \sin \theta _{D-1},
\end{equation*}%
\begin{equation*}
\vdots
\end{equation*}%
\begin{equation*}
x_{j}=r\cos \theta _{j-1}\sin \theta _{j}\cdots \sin \theta _{D-1},\text{ }%
3\leq j\leq D-1,
\end{equation*}%
\begin{equation*}
\vdots
\end{equation*}%
\begin{equation*}
x_{D-1}=r\cos \theta _{D-2}\sin \theta _{D-1},
\end{equation*}%
\begin{equation}
x_{D}=r\cos \theta _{D-1},\text{ }\sum\limits_{j=1}^{D}x_{j}^{2}=r^{2},D\geq
2.
\end{equation}%
We have $x_{1}=r\cos \theta _{1},$ $x_{2}=r\sin \theta _{1},$ $\theta
_{1}=\varphi ,$ for $D=2$ and $x_{1}=r\cos \theta _{1},$ $x_{2}=r\sin \theta
_{1}\sin \theta _{2},$ $x_{3}=r\cos \theta _{2},$ $\theta _{2}=\theta ,$ for
$D=3.$ The volume element of the configuration space is given by%
\begin{equation}
\prod\limits_{j=1}^{D}dx_{j}=r^{D-1}d^{3}rd\Omega ,\text{ }d\Omega
=\prod\limits_{j=1}^{D-1}(\sin \theta _{j})^{j-1}d\theta _{j},
\end{equation}%
where $r\in \lbrack 0,\infty ),$ $\theta _{1}\in \lbrack 0,2\pi ]$ and $%
\theta _{j}\in \lbrack 0,\pi ],$ $j\in \lbrack 2,D-1].$

The wave function $\psi _{l_{1}\cdots l_{D-2}}^{(l_{D-1}=l)}(\mathbf{x})$
with a given angular momentum $l$ can be decomposed as a product of a radial
wave function $R_{l}(r)$ and the normalized hyper-spherical harmonics $%
Y_{l_{1}\cdots l_{D-2}}^{(l)}(\theta _{1},\theta _{2},\cdots ,\theta _{D-1})$
as%
\begin{equation}
\psi _{l_{1}\cdots l_{D-2}}^{(l_{D-1}=l)}(\mathbf{x})=R_{l}(r)Y_{l_{1}\cdots
l_{D-2}}^{(l)}(\theta _{1},\theta _{2},\cdots ,\theta _{D-1}),
\end{equation}%
where%
\begin{equation}
Y_{l_{1}\cdots l_{D-2}}^{(l)}(\theta _{1},\theta _{2},\cdots ,\theta
_{D-1})=H(\theta _{1})H(\theta _{2},\cdots ,\theta _{D-2})H(\theta _{D-1}),
\end{equation}%
which is the simultaneous eigenfunction of $L_{j}^{2}:$%
\begin{equation*}
L_{j}^{2}Y_{l_{1}\cdots l_{D-2}}^{(l)}(\theta _{1},\theta _{2},\cdots
,\theta _{D-1})=l_{j}(l_{j}+j-1)Y_{l_{1}\cdots l_{D-2}}^{(l)}(\theta
_{1},\theta _{2},\cdots ,\theta _{D-1}),
\end{equation*}%
\begin{equation*}
l=0,1,\cdots ,l_{k}=0,1,\cdots ,l_{k+1},\text{ }j\in \lbrack 1,D-1],\text{ }%
k\in \lbrack 2,D-2],
\end{equation*}%
\begin{equation*}
l_{1}=-l_{2},-l_{2}+1,\cdots ,l_{2}-1,l_{2},
\end{equation*}%
\begin{equation}
L_{D-1}^{2}Y_{l_{1}\cdots l_{D-2}}^{(l)}(\theta _{1},\theta _{2},\cdots
,\theta _{D-1})=l(l+D-2)Y_{l_{1}\cdots l_{D-2}}^{(l)}(\theta _{1},\theta
_{2},\cdots ,\theta _{D-1}).
\end{equation}%
The angular momentum operators $L_{j}^{2}$ are defined as%
\begin{equation*}
L_{1}^{2}=-\frac{\partial ^{2}}{\partial \theta _{1}^{2}},
\end{equation*}%
\begin{equation*}
L_{k}^{2}=\sum\limits_{a<b=2}^{k+1}L_{ab}^{2}=-\frac{1}{\sin ^{k-1}\theta
_{k}}\frac{\partial }{\partial \theta _{k}}\left( \sin ^{k-1}\theta _{k}%
\frac{\partial }{\partial \theta _{k}}\right) +\frac{L_{k-1}^{2}}{\sin
^{2}\theta _{k}},\text{ }2\leq k\leq D-1,
\end{equation*}%
\begin{equation}
L_{ab}=-i\left[ x_{a}\frac{\partial }{\partial x_{b}}-x_{b}\frac{\partial }{%
\partial x_{a}}\right] .
\end{equation}%
Employing the method of separation of variables and substituting Eqs.
(5)-(7) into Eq. (2), we obtain the following Schr\"{o}dinger-like equation:%
\begin{equation}
R_{l}{}^{\prime \prime }(r)+\frac{(D-1)}{r}R_{l}^{\prime }(r)+\left\{ \frac{%
l(l+D-2)}{r^{2}}+\left[ E_{nl}-V(r)\right] ^{2}-\left[ m_{0}+S(r)\right]
^{2}\right\} R_{l}(r)=0,
\end{equation}%
where $l(l+D-2)/r^{2}$ is known as the centrifugal term. Furthermore, using $%
R_{l}(r)=r^{-(D-1)/2}g(r),$ it is straightforward to find out the radial
wave equation:%
\begin{equation}
g^{\prime \prime }(r)+\left\{ \left[ E_{nl}-V(r)\right] ^{2}-\left[
m_{0}+S(r)\right] ^{2}-\frac{(D+2l-1)(D+2l-3)}{4r^{2}}\right\} g(r)=0.
\end{equation}%
Obviously, Eq. (10) can be analytically solved [17,27,30] only for $l=0$ ($s$%
-wave) case. For $l\neq 0$ case$,$ we have to use an approximation for the
centrifugal term similar to the non-relativistic cases which is valid for $%
q=1$ value [29,32,33,38-43]:

\begin{equation}
\frac{1}{r^{2}}\approx \frac{\alpha ^{2}e^{-\alpha r}}{\left( 1-qe^{-\alpha
r}\right) ^{2}}.\text{ }
\end{equation}%
We follow the model used by Qiang \textit{et al}. [32] and rewrite Eq. (10)
for the scalar and vector general Hulth\'{e}n-type potentials (1) as,%
\begin{equation*}
g^{\prime \prime }(r)+\left[ E_{nl}^{2}-m_{0}^{2}+\frac{2\left(
m_{0}S_{0}+E_{nl}V_{0}\right) e^{-\alpha r}}{1-qe^{-\alpha r}}-\frac{\left(
S_{0}^{2}-V_{0}^{2}\right) e^{-2\alpha r}}{\left( 1-qe^{-\alpha r}\right)
^{2}}-\frac{(D+2l-1)(D+2l-3)}{4r^{2}}\right]
\end{equation*}%
\begin{equation}
\times g(r)=0.
\end{equation}%
On the other hand, the other angular equations, obtained from the separation
procedures, are [21,22]:%
\begin{equation*}
\left[ \frac{1}{\sin ^{j-1}\theta _{j}}\frac{d}{d\theta _{j}}\left( \sin
^{j-1}\theta _{j}\frac{d}{d\theta _{j}}\right) +l_{j}(l_{j}+j-1)-\frac{%
l_{j-1}(l_{j-1}+j-2)}{\sin ^{2}\theta _{j}}\right] H(\theta _{2},\cdots
,\theta _{D-2})=0,
\end{equation*}%
\begin{equation}
j\in \lbrack 2,D-2],
\end{equation}%
\begin{equation}
\left[ \frac{1}{\sin ^{D-2}\theta _{D-1}}\frac{d}{d\theta _{D-1}}\left( \sin
^{D-2}\theta _{D-1}\frac{d}{d\theta _{D-1}}\right) +l(l+D-2)-\frac{%
L_{D-2}^{2}}{\sin ^{2}\theta _{D-1}}\right] H(\theta _{D-1})=0,
\end{equation}%
\begin{equation}
\frac{d^{2}H((\theta _{1})}{d\theta _{1}^{2}}+l_{1}^{2}H((\theta _{1})=0.
\end{equation}%
It is well-known that the solution of Eq. (15) is%
\begin{equation}
H_{l_{1}}(\theta _{1})=\frac{1}{\sqrt{2\pi }}\exp (\pm il_{1}\theta _{1}),%
\text{ \ }l_{1}=0,1,2,\cdots .
\end{equation}

Hence, the above Eqs. (12)-(14), have to be solved by using N-U method
[20-31,36] which is reviewed briefly in the following Section.

\section{Nikiforov-Uvarov Method}

The N-U method is breifly outlined here and the details can be found in
[20-31,36,37,44]. N-U method is proposed to solve the second-order
differential equation of the hypergeometric type:%
\begin{equation}
\psi _{n}^{\prime \prime }(s)+\frac{\widetilde{\tau }(s)}{\sigma (s)}\psi
_{n}^{\prime }(s)+\frac{\widetilde{\sigma }(s)}{\sigma ^{2}(s)}\psi
_{n}(s)=0,
\end{equation}%
where $\sigma (s)$ and $\widetilde{\sigma }(s)$ are polynomials, at most of
second-degree, and $\widetilde{\tau }(s)$ is a first-degree polynomial.
Using a wave function, $\psi _{n}(s),$ of \ the simple form%
\begin{equation}
\psi _{n}(s)=\phi _{n}(s)y_{n}(s),
\end{equation}%
reduces Eq. (17) into an equation of a hypergeometric type

\begin{equation}
\sigma (s)y_{n}^{\prime \prime }(s)+\tau (s)y_{n}^{\prime }(s)+\lambda
y_{n}(s)=0,
\end{equation}%
where%
\begin{equation}
\sigma (s)=\pi (s)\frac{\phi (s)}{\phi ^{\prime }(s)},
\end{equation}

\begin{equation}
\tau (s)=\widetilde{\tau }(s)+2\pi (s),\text{ }\tau ^{\prime }(s)<0,
\end{equation}%
and $\lambda $ is a parameter defined as%
\begin{equation}
\lambda =\lambda _{n}=-n\tau ^{\prime }(s)-\frac{n\left( n-1\right) }{2}%
\sigma ^{\prime \prime }(s),\text{ \ \ \ \ \ \ }n=0,1,2,\cdots .
\end{equation}%
The polynomial $\tau (s)$ with the parameter $s$ and prime factors show the
differentials at first degree be negative. The other part $y_{n}(s)$ of the
wave function Eq. (18) is the hypergeometric-type function whose polynomial
solutions are given by Rodrigues relation%
\begin{equation}
y_{n}(s)=\frac{B_{n}}{\rho (s)}\frac{d^{n}}{ds^{n}}\left[ \sigma ^{n}(s)\rho
(s)\right] ,
\end{equation}%
where $B_{n}$ is the normalization constant and the weight function $\rho
(s) $ can be found by [44]%
\begin{equation}
\omega ^{\prime }(s)-\frac{\tau (s)}{\sigma (s)}\omega (s)=0,\text{ }\omega
(s)=\sigma (s)\rho (s).
\end{equation}%
The function $\pi (s)$ and the parameter $\lambda $ are defined as%
\begin{equation}
\pi (s)=\frac{\sigma ^{\prime }(s)-\widetilde{\tau }(s)}{2}\pm \sqrt{\left(
\frac{\sigma ^{\prime }(s)-\widetilde{\tau }(s)}{2}\right) ^{2}-\widetilde{%
\sigma }(s)+k\sigma (s)},
\end{equation}%
\begin{equation}
\lambda =k+\pi ^{\prime }(s),
\end{equation}%
where $\pi (s)$ has to be a polynomial of degree at most one. To obtain $k,$
the expression under the square root sign in Eq. (25) can be arranged to be
the square of a polynomial of first degree [44]. This is possible only if
its discriminant is zero. Finally, the energy eigenvalues are obtained by
comparing Eqs. (22) and (26).

\section{Solutions of the Radial and Angle-Dependent Equations}

\subsection{The $D$-dimensional angular equations}

At the beginning, we rewrite Eqs. (13) and (14) representing the angular
wave equations in the following simple forms [21,22] :%
\begin{equation}
\frac{d^{2}H(\theta _{j})}{d\theta _{j}^{2}}+(j-1)ctg\theta _{j}\frac{%
dH(\theta _{j})}{d\theta _{j}}+\left( \Lambda _{j}-\frac{\Lambda _{j-1}}{%
\sin ^{2}\theta _{j}}\right) H(\theta _{j})=0,\text{ }j\in \lbrack
2,D-2],D>3,
\end{equation}%
\begin{equation}
\frac{d^{2}H(\theta _{D-1})}{d\theta _{D-1}^{2}}+(D-2)ctg\theta _{D-1}\frac{%
dH(\theta _{D-1})}{d\theta _{D-1}}+\left[ l(l+D-2)-\frac{\Lambda _{D-2}}{%
\sin ^{2}\theta _{D-1}}\right] H(\theta _{D-1})=0,
\end{equation}%
where $\Lambda _{p}=l_{p}(l_{p}+p-1),$ $p=j-1,$ $j$ with $l_{p}=0,1,2,\cdots
$ are the angular quantum numbers in $D$-dimensions. Employing $s=\cos
\theta _{j},$ we transform Eq. (27) to the associated-Legendre equation%
\begin{equation}
\frac{d^{2}H(s)}{ds^{2}}-\frac{js}{1-s^{2}}\frac{dH(s)}{ds}+\frac{\Lambda
_{j}-\Lambda _{j-1}-\Lambda _{j}s^{2}}{(1-s^{2})^{2}}H(s)=0,\text{ }j\in
\lbrack 2,D-2],D>3.
\end{equation}%
By comparing Eqs. (29) and (17), the corresponding polynomials are obtained%
\begin{equation}
\widetilde{\tau }(s)=-js,\ \ \ \sigma (s)=1-s^{2},\ \ \widetilde{\sigma }%
(s)=-\Lambda _{j}s^{2}+\Lambda _{j}-\Lambda _{j-1}.
\end{equation}%
Inserting the above expressions into Eq. (25) and taking $\sigma ^{\prime
}(s)=-2s$, one obtains the following function:%
\begin{equation}
\pi (s)=\frac{(j-2)}{2}s\pm \sqrt{\left[ \left( \frac{j-2}{2}\right)
^{2}+\Lambda _{j}-k\right] s^{2}+k-\Lambda _{j}+\Lambda _{j-1}}.
\end{equation}%
Following the method, the polynomial $\pi (s)$ is found to have the
following four possible values:%
\begin{equation}
\pi (s)=\left\{
\begin{array}{cc}
\left( \frac{j-2}{2}+\widetilde{\Lambda }_{j-1}\right) s & \text{\ for }%
k_{1}=\Lambda _{j}-\Lambda _{j-1}, \\
\left( \frac{j-2}{2}-\widetilde{\Lambda }_{j-1}\right) s & \text{\ for }%
k_{1}=\Lambda _{j}-\Lambda _{j-1}, \\
\frac{(j-2)}{2}s+\widetilde{\Lambda }_{j-1} & \text{\ for }k_{2}=\Lambda
_{j}+\left( \frac{j-2}{2}\right) ^{2}, \\
\frac{(j-2)}{2}s-\widetilde{\Lambda }_{j-1} & \text{\ for }k_{2}=\Lambda
_{j}+\left( \frac{j-2}{2}\right) ^{2},%
\end{array}%
\right.
\end{equation}%
where $\widetilde{\Lambda }_{p}=l_{p}+(p-1)/2,$ with $p=j-1,$ $j$ and $j\in
\lbrack 2,D-2],$ $D>3.$ Imposing the condition $\tau ^{\prime }(s)<0,$ where
$\tau ^{\prime }=\tau ^{\prime }(l_{j-1}),$ we may select the following
physical solutions:%
\begin{equation}
k_{1}=\Lambda _{j}-\Lambda _{j-1}\text{\ \ and \ \ }\pi (s)=-l_{j-1}s,\text{
}j\in \lbrack 2,D-2],D>3,
\end{equation}%
which yields%
\begin{equation}
\tau (s)=-2(1+\widetilde{\Lambda }_{j-1})s.
\end{equation}%
Using Eqs. (22) and (26), give%
\begin{equation}
\lambda _{n_{j}}=2n_{j}(1+\widetilde{\Lambda }_{j-1})+n_{j}(n_{j}-1),
\end{equation}%
\begin{equation}
\lambda =\Lambda _{j}-\Lambda _{j-1}-l_{j-1},
\end{equation}%
and setting $\lambda =\lambda _{n_{j}},$ we obtain%
\begin{equation}
n_{j}=l_{j}-l_{j-1}.
\end{equation}%
Substituting Eqs. (30), (33) and (34) into Eqs. (20)-(21) and (23)-(24), give%
\begin{equation}
\phi (s)=\left( 1-s^{2}\right) ^{l_{j-1}/2},\text{ }\rho (s)=\left(
1-s^{2}\right) ^{l_{j-1}+(j-2)/2}.
\end{equation}%
Further, substituting the weight function $\rho (s)$ given in Eq. (38) into
the Rodrigues relation (23) gives%
\begin{equation}
y_{n_{j}}(s)=A_{n_{j}}\left( 1-s^{2}\right) ^{-\widetilde{\Lambda }_{j-1}}%
\frac{d^{n_{j}}}{ds^{n_{j}}}\left( 1-s^{2}\right) ^{n_{j}+\widetilde{\Lambda
}_{j-1}},
\end{equation}%
where $A_{n_{j}}$ is the normalization factor. Finally, the angular wave
functions are%
\begin{equation}
H_{n_{j}}(\theta _{j})=N_{n_{j}}\left( \sin \theta _{j}\right)
^{^{l_{j}-n_{j}}}P_{n_{j}}^{(l_{j}-n_{j}+(j-2)/2,l_{j}-n_{j}+(j-2)/2)}(\cos
\theta _{j}),
\end{equation}%
where the quantum numbers $n_{j}$ are defined in Eq. (37) and the
normalization factor is%
\begin{equation}
N_{n_{j}}=\sqrt{\frac{\left( 2l_{j}+j-1\right) n_{j}!}{2\Gamma \left(
l_{j}+l_{j-1}+j-2\right) }},\text{ \ }j\in \lbrack 2,D-2],\text{ }D>3.
\end{equation}%
Likewise, in solving Eq. (28), we introduce a new variable $s=\cos \theta
_{D-1}.$ Thus, we can also rearrange it as the universal associated-Legendre
differential equation%
\begin{equation}
\frac{d^{2}H(s)}{ds^{2}}-\frac{(D-1)s}{1-s^{2}}\frac{dH(s)}{ds}+\frac{\nu
(1-s^{2})-\Lambda _{D-2}}{(1-s^{2})^{2}}H(s)=0,
\end{equation}%
where%
\begin{equation}
\nu =l(l+D-2).
\end{equation}%
Equation (42) has been recently solved in $2D$ and $3D$ by the N-U method in
[45,46]. By comparing Eqs. (42) and (17), the corresponding polynomials are
obtained%
\begin{equation}
\widetilde{\tau }(s)=-(D-1)s,\text{ \ \ \ }\sigma (s)=1-s^{2},\text{ \ \ }%
\widetilde{\sigma }(s)=-\nu s^{2}+\nu -\Lambda _{D-2}.
\end{equation}%
Inserting the above expressions into Eq. (25) and taking $\sigma ^{\prime
}(s)=-2s$, one obtains the following function:%
\begin{equation}
\pi (s)=\frac{(D-3)}{2}s\pm \sqrt{\left[ \left( \frac{D-3}{2}\right)
^{2}+\nu -k\right] s^{2}+k-\nu +\Lambda _{D-2}},
\end{equation}%
which gives the possible solutions:
\begin{equation}
\pi (s)=\left\{
\begin{array}{cc}
\left( \frac{D-3}{2}+m^{\prime }\right) s & \text{\ for }k_{1}=\nu -\Lambda
_{D-2}, \\
\left( \frac{D-3}{2}-m^{\prime }\right) s & \text{\ for }k_{1}=\nu -\Lambda
_{D-2}, \\
\frac{(D-3)}{2}s+m^{\prime } & \text{\ for }k_{2}=\nu +\left( \frac{D-3}{2}%
\right) ^{2}, \\
\frac{(D-3)}{2}s-m^{\prime } & \text{\ for }k_{2}=\nu +\left( \frac{D-3}{2}%
\right) ^{2},%
\end{array}%
\right.
\end{equation}%
where $m^{\prime }=l_{D-2}+\frac{D-3}{2}.$ Imposing the condition $\tau
^{\prime }(s)<0,$ where $\tau ^{\prime }=\tau ^{\prime }(l_{D-2}),$ one may
select:
\begin{equation}
k_{1}=\nu -\Lambda _{D-2}\text{ \ \ and \ \ }\pi (s)=-l_{D-2}s,
\end{equation}%
which yields
\begin{equation}
\tau (s)=-2(1+m^{\prime })s.
\end{equation}%
The following expressions for $\lambda $ are obtained, respectively,

\begin{equation}
\lambda =\lambda _{n_{D-1}}=2n_{D-1}(1+m^{\prime })+n_{D-1}(n_{D-1}-1),
\end{equation}%
\begin{equation}
\lambda =l(l+D-2)-l_{D-2}(l_{D-2}+D-2),
\end{equation}%
and from $\lambda =\lambda _{n_{D-1}},$ we obtain the angular momentum
quantum number:%
\begin{equation}
l=l_{D-2}+n_{D-1}.
\end{equation}%
where $n_{D-1}=n$ and $l_{1}=m$ in $3D$ space [46]. Using Eqs. (20)-(21) and
(23)-(24), we obtain%
\begin{equation}
\phi (s)=\left( 1-s^{2}\right) ^{l_{D-2}/2},\text{ }\rho (s)=\left(
1-s^{2}\right) ^{m^{\prime }}.
\end{equation}%
The Rodrigues relation (23) gives the following wave functions:%
\begin{equation}
y_{n_{D-1}}(s)=B_{n_{D-1}}\left( 1-s^{2}\right) ^{-m^{\prime }}\frac{%
d^{n_{D-1}}}{ds^{n_{D-1}}}\left( 1-s^{2}\right) ^{n_{D-1}+m^{\prime }},
\end{equation}%
where $B_{n_{D-1}}$ is the normaliation factor. Finally, the angular wave
functions are%
\begin{equation}
H_{n_{D-1}}(\theta _{D-1})=N_{n_{D-1}}\left( \sin \theta _{D-1}\right)
^{l_{D-2}}P_{n_{D-1}}^{(m^{\prime },m^{\prime })}(\cos \theta _{D-1}),
\end{equation}%
where the normalization factor is%
\begin{equation}
N_{n_{D-1}}=\sqrt{\frac{\left( 2n_{D-1}+2m^{\prime }+1\right) n_{D-1}!}{%
2\Gamma \left( n_{D-1}+2m^{\prime }\right) }},
\end{equation}%
with $m^{\prime }$ is defined after Eq. (46).

\subsection{Bound States of the $D$-dimensional Radial Equation}

The $D$-dimensional Klein-Gordon radial energy eigenvalue equation with the
vector and scalar general Hulth\'{e}n-type potentials can be rewritten as
\begin{equation*}
g^{\prime \prime }(r)+\left[ E_{nl}^{2}-m_{0}^{2}+\frac{2\left(
m_{0}S_{0}+E_{nl}V_{0}\right) e^{-\alpha r}}{1-qe^{-\alpha r}}\right.
\end{equation*}%
\begin{equation}
-\left. \frac{\frac{1}{4}(D+2l-1)(D+2l-3)\alpha ^{2}e^{-\alpha r}+\left(
S_{0}^{2}-V_{0}^{2}\right) e^{-2\alpha r}}{\left( 1-qe^{-\alpha r}\right)
^{2}}\right] g(r),\text{ }g(0)=0,
\end{equation}%
On account of the wave function $g(r)$ satisfying the standard bound state
condition, which is when $r\rightarrow \infty $, the wave function $%
g(r)\rightarrow 0.$ Equation (56) can be further transformed by using a new
variable $s=qe^{-\alpha r}$ $(r\in \lbrack 0,\infty ),$ $s\in \lbrack q,0]),$%
\begin{equation}
g^{\prime \prime }(s)+\frac{(1-s)}{s(1-s)}g^{\prime }(s)+\frac{1}{\left[
s(1-s)\right] ^{2}}\left[ -\varepsilon _{nl}^{2}+(\beta _{1}-\gamma
+2\varepsilon _{nl}^{2})s-(\beta _{1}+\beta _{2}+\varepsilon _{nl}^{2})s^{2}%
\right] g(s)=0,
\end{equation}%
where%
\begin{equation}
\varepsilon _{nl}=\frac{\sqrt{m_{0}^{2}-E_{nl}^{2}}}{\alpha },\text{ \ }%
\beta _{1}=\frac{2\left( m_{0}S_{0}+E_{nl}V_{0}\right) }{\alpha ^{2}q},\text{
}\beta _{2}=\frac{S_{0}^{2}-V_{0}^{2}}{\alpha ^{2}q^{2}},\text{ }\gamma =%
\frac{(D+2l-1)(D+2l-3)}{4q}.
\end{equation}%
For bound states, $E_{nl}\leq m_{0},$ $\varepsilon _{nl}\geq 0$ [17,47,48].
Comparing Eqs. (57) and (17), the corresponding polynomials are obtained:%
\begin{equation}
\widetilde{\tau }(s)=1-s,\sigma (s)=s(1-s),\widetilde{\sigma }%
(r)=-\varepsilon _{nl}^{2}+(\beta _{1}-\gamma +2\varepsilon
_{nl}^{2})s-(\beta _{1}+\beta _{2}+\varepsilon _{nl}^{2})s^{2}.
\end{equation}%
The substitution of Eq. (59) into Eq. (25) and taking $\sigma ^{\prime
}(s)=1-2s,$ give the polynomial:%
\begin{equation}
\pi (s)=-\frac{s}{2}\pm \frac{1}{2}\sqrt{\left[ 1+4(\beta _{1}+\beta
_{2}+\varepsilon _{nl}^{2}-k)\right] s^{2}+4(k-\beta _{1}+\gamma
-2\varepsilon _{nl}^{2})s+4\varepsilon _{nl}^{2}}.
\end{equation}%
If the expression under the square root in Eq. (60) is set equal to zero and
solved for $k,$ we obtain:%
\begin{equation}
k=\beta _{1}-\gamma \pm \varepsilon _{nl}\sqrt{1+4(\beta _{2}+\gamma )}.
\end{equation}%
In view of that, we arrive at the following four possible functions of $\pi
(s):$%
\begin{equation}
\pi (s)=\left\{
\begin{array}{cc}
-\frac{s}{2}+\varepsilon _{nl}-\left[ \varepsilon _{nl}-\frac{1}{2}\sqrt{%
1+4(\beta _{2}+\gamma )}\right] s & \text{\ for }k_{1}=\beta _{1}-\gamma
+\varepsilon _{nl}\sqrt{1+4(\beta _{2}+\gamma )}, \\
-\frac{s}{2}-\varepsilon _{nl}+\left[ \varepsilon _{nl}-\frac{1}{2}\sqrt{%
1+4(\beta _{2}+\gamma )}\right] s & \text{\ for }k_{1}=\beta _{1}-\gamma
+\varepsilon _{nl}\sqrt{1+4(\beta _{2}+\gamma )}, \\
-\frac{s}{2}+\varepsilon _{nl}-\left[ \varepsilon _{nl}+\frac{1}{2}\sqrt{%
1+4(\beta _{2}+\gamma )}\right] s & \text{\ for }k_{2}=\beta _{1}-\gamma
-\varepsilon _{nl}\sqrt{1+4(\beta _{2}+\gamma )}, \\
-\frac{s}{2}-\varepsilon _{nl}+\left[ \varepsilon _{nl}+\frac{1}{2}\sqrt{%
1+4(\beta _{2}+\gamma )}\right] s & \text{\ for }k_{2}=\beta _{1}-\gamma
-\varepsilon _{nl}\sqrt{1+4(\beta _{2}+\gamma )}.%
\end{array}%
\right.
\end{equation}%
The correct value of $\pi (s)$ is chosen such that the function $\tau (s)$
in Eq. (21) must have negative derivative [44]. So we can select the
physical values to be%
\begin{equation}
k=\beta _{1}-\gamma -\varepsilon _{nl}\sqrt{1+4(\beta _{2}+\gamma )}\text{ \
\ and \ \ }\pi (s)=-\frac{s}{2}+\varepsilon _{nl}-\left[ \varepsilon _{nl}+%
\frac{1}{2}\sqrt{1+4(\beta _{2}+\gamma )}\right] s,
\end{equation}%
which yield%
\begin{equation*}
\tau (s)=1+2\varepsilon _{nl}-2\left[ 1+\varepsilon _{nl}+\frac{1}{2}\sqrt{%
1+4(\beta _{2}+\gamma )}\right] s,
\end{equation*}%
\begin{equation}
\tau ^{\prime }(s)=-2\left[ 1+\varepsilon _{nl}+\frac{1}{2}\sqrt{1+4(\beta
_{2}+\gamma )}\right] <0.
\end{equation}%
Using Eqs. (22) and (26), the following expressions for $\lambda $ are
obtained, respectively,%
\begin{equation}
\lambda =\lambda _{n}=n^{2}+\left[ 1+2\varepsilon _{nl}+\sqrt{1+4(\beta
_{2}+\gamma )}\right] n,\text{ }(n=0,1,2,\cdots ),
\end{equation}%
\begin{equation}
\lambda =\beta _{1}-\gamma -\frac{1}{2}(1+2\varepsilon _{nl})\left[ 1+\sqrt{%
1+4(\beta _{2}+\gamma )}\right] ,
\end{equation}%
where $n$ is the radial quantum number. Solving the last two equations and
using $\beta _{2}=\delta _{\pm }^{2}-\delta _{\pm }-\gamma ,$ give%
\begin{equation*}
\varepsilon _{nl}^{(D)}=\frac{\left( \beta _{1}-\gamma -n^{2}\right)
-(2n+1)\delta _{\pm }}{2(n+\delta _{\pm })}=\frac{4q\left[ \beta
_{1}-n^{2}-(2n+1)\delta _{\pm }\right] -(D+2l-1)(D+2l-3)}{8q(n+\delta _{\pm
})}
\end{equation*}%
\begin{equation*}
=\frac{2q(m_{0}S_{0}+E_{nl}V_{0})+S_{0}^{2}-V_{0}^{2}}{2q^{2}\alpha
^{2}(n+\delta _{\pm })}-\frac{n+\delta _{\pm }}{2},\text{ }(n=0,1,2,\cdots ),
\end{equation*}%
\begin{equation}
\delta _{\pm }=\frac{1}{2}\left( 1\pm \frac{a}{q}\right) ,\text{ }a=\sqrt{%
q^{2}+\frac{4(S_{0}^{2}-V_{0}^{2})}{\alpha ^{2}}+q(D+2l-1)(D+2l-3)},\text{ }%
(l=0,1,2,\cdots ),
\end{equation}%
with $\delta =\delta _{+}$ for $q>0$ and $\delta =\delta _{-}$ for $q<0.$
Furthermore, from setting $\varepsilon _{nl}^{(D)}=\varepsilon _{nl}$ and
using Eqs. (58) and (67)$,$ we obtain the energy equation%
\begin{equation}
\sqrt{m_{0}^{2}-E_{nl}^{2}}=\frac{%
2q(m_{0}S_{0}+E_{nl}V_{0})+S_{0}^{2}-V_{0}^{2}}{2q^{2}\alpha (n+\delta )}-%
\frac{\alpha (n+\delta )}{2},
\end{equation}%
or equivalently the explicit expression of the energy eigenvalues%
\begin{equation*}
E_{nl}^{(D)}=\frac{\eta _{nl}V_{0}}{2q}\pm \kappa _{nl}\sqrt{\frac{m_{0}^{2}%
}{4V_{0}^{2}+\kappa _{nl}^{2}}-\left( \frac{\eta _{nl}}{4q}\right) ^{2}}%
,16q^{2}m_{0}^{2}\geq \eta _{nl}^{2}\left( 4V_{0}^{2}+\kappa
_{nl}^{2}\right) ,
\end{equation*}%
\begin{equation*}
\eta _{nl}=\frac{4(V_{0}^{2}-S_{0}^{2})+\kappa _{nl}^{2}-8qm_{0}S_{0}}{%
4V_{0}^{2}+\kappa _{nl}^{2}},
\end{equation*}%
\begin{equation*}
\kappa _{nl}=q\alpha (2n+1)\pm \sqrt{q^{2}\alpha
^{2}+4(S_{0}^{2}-V_{0}^{2})+q\alpha ^{2}(D+2l-1)(D+2l-3)},
\end{equation*}%
\begin{equation}
\text{(}n=0,1,2,3,\cdots ,\text{ }l=0,1,2,\cdots ),
\end{equation}%
and
\begin{equation}
q^{2}\alpha ^{2}+q\alpha ^{2}(D+2l-2)^{2}+4S_{0}^{2}\geq q\alpha
^{2}+4V_{0}^{2},
\end{equation}%
is a constraint over the potential parameters. In the case of pure vector
potential ($S_{0}=0,V_{0}\neq 0)$, the energy equation (69) reduces to
\begin{equation*}
E_{nl}^{(D)}=\frac{V_{0}}{2q}\pm \kappa _{nl}\sqrt{\frac{m_{0}^{2}}{%
4V_{0}^{2}+\kappa _{nl}^{2}}-\left( \frac{1}{4q}\right) ^{2}},\text{ }%
16q^{2}m_{0}^{2}\geq 4V_{0}^{2}+\kappa _{nl}^{2},
\end{equation*}%
\begin{equation*}
\kappa _{nl}=q\alpha (2n+1)\pm \sqrt{q^{2}\alpha ^{2}+q\alpha
^{2}(D+2l-1)(D+2l-3)-4V_{0}^{2}},\text{ }D\geq 1,
\end{equation*}%
\begin{equation}
\text{(}n=0,1,2,3,\cdots ,\text{ }l=0,1,2,\cdots ),
\end{equation}%
with the constraint
\begin{equation}
q^{2}\alpha ^{2}+q\alpha ^{2}(D+2l-2)^{2}\geq q\alpha ^{2}+4V_{0}^{2},
\end{equation}%
for real bound states. From the above result, it is not difficult to
conclude that the two energy solutions are valid for the particle and the
second one corresponds to the anti-particle energy.

The restriction that gives the critical coupling value leads to the result%
\begin{equation}
n\leq \frac{1}{q\alpha }\left( \sqrt{4q^{2}m_{0}^{2}-V_{0}^{2}}-\sqrt{\frac{%
q\alpha ^{2}}{4}\left[ q+(D+2l-1)(D+2l-3)\right] -V_{0}^{2}}\right) -\frac{1%
}{2},
\end{equation}%
i.e., there are only finitly many eigenvalues. In order that at least one
level might exist, it is necessary that the inequality:%
\begin{equation}
q\alpha +\sqrt{q\alpha ^{2}\left[ q+(D+2l-1)(D+2l-3)\right] -4V_{0}^{2}}\leq
2\sqrt{4q^{2}m_{0}^{2}-V_{0}^{2}},
\end{equation}%
is fulfilled. As can be seen from \ Eq. (73), there are at most only two
lower-lying states $(n=0,1)$ for the Klein-Gordon particle of mass unity
when the parameter $\alpha =1$ and $q=\pm 1$ for any arbitrary value of $%
V_{0}.$ Therefore, we have the inequality:%
\begin{equation}
n\leq \pm \left( \sqrt{4-V_{0}^{2}}-\sqrt{\left( \frac{D+2l-2}{2}\right)
^{2}-V_{0}^{2}}\right) -\frac{1}{2}.
\end{equation}%
For instance, if one selects $V_{0}=(D+2l-2)/2,$ then the real bound states
must be restricted by $n\leq \left( \sqrt{16-(D+2l-2)^{2}}-1\right) /2.$

Having solved the $D$-dimensional Klein-Gordon equation for scalar and
vector general Hulth\'{e}n-type potentials, we should make some remarks.

(i) For $s$-wave ($l=0$), the exact energy eigenvalues of the $1D$
Klein-Gordon equation becomes%
\begin{equation}
E_{n}=\frac{\eta _{n}V_{0}}{2q}\pm \frac{\kappa _{n}}{4q\left(
4V_{0}^{2}+\kappa _{n}^{2}\right) }\sqrt{\left[ \kappa _{n}^{2}+4\left(
V_{0}^{2}-S_{0}^{2}\right) \right] \left[ \left( 2S_{0}+4qm_{0}\right)
^{2}-\kappa _{n}^{2}-4V_{0}^{2}\right] },
\end{equation}%
where%
\begin{equation*}
\kappa _{n}=q\alpha (2n+1)+\sqrt{q^{2}\alpha ^{2}+4(S_{0}^{2}-V_{0}^{2})},
\end{equation*}%
\begin{equation}
\eta _{n}=\frac{4(V_{0}^{2}-S_{0}^{2})+\kappa _{n}^{2}-8qm_{0}S_{0}}{%
4V_{0}^{2}+\kappa _{n}^{2}}.
\end{equation}%
In order that at least one level might exist, it is necessary that the
inequality%
\begin{equation}
16q^{2}m_{0}^{2}\geq \eta _{n}^{2}\left( 4V_{0}^{2}+\kappa _{n}^{2}\right) ,%
\text{ }q^{2}\alpha ^{2}+4S_{0}^{2}\geq 4V_{0}^{2},\text{ }
\end{equation}%
is fulfilled. In the case of pure vector potential ($S_{0}=0,V_{0}\neq 0),$
the energy spectrum
\begin{equation}
E_{n}=\frac{V_{0}}{2q}\pm \left[ q\alpha (2n+1)+\sqrt{q^{2}\alpha
^{2}-4V_{0}^{2}}\right] \sqrt{\frac{m_{0}^{2}}{4V_{0}^{2}+\left[ q\alpha
(2n+1)+\sqrt{q^{2}\alpha ^{2}-4V_{0}^{2}}\right] ^{2}}-\frac{1}{16q^{2}}},
\end{equation}%
with%
\begin{equation}
16q^{2}m_{0}^{2}\geq 4V_{0}^{2}+\left[ q\alpha (2n+1)+\sqrt{q^{2}\alpha
^{2}-4V_{0}^{2}}\right] ^{2},\text{ }q\alpha \geq 2V_{0}.
\end{equation}%
We notice that the result given in Eq. (79) is identical to Eq. (31) of Ref.
[30]. There are only two lower-lying states $(n=0,1)$ for the Klein-Gordon
particle of a rest mass $m_{0}=1.0$ and $\alpha =1.0$. For example, one may
calculate the ground state energy for the coupling strength $V_{0}=q\alpha
/2 $ as
\begin{equation}
E_{0}=\frac{V_{0}}{2q}\left[ 1\pm \sqrt{\frac{2m_{0}q^{2}}{V_{0}^{2}}-1}%
\right] .
\end{equation}%
Further, in the case of pure scalar potential ($V_{0}=0,S_{0}\neq 0),$ the
energy spectrum%
\begin{equation*}
E_{n}=\pm \frac{1}{4q\left[ q\alpha (2n+1)+\sqrt{q^{2}\alpha ^{2}+4S_{0}^{2}}%
\right] }\sqrt{\left[ q\alpha (2n+1)+\sqrt{q^{2}\alpha ^{2}+4S_{0}^{2}}%
\right] ^{2}-4S_{0}^{2}}
\end{equation*}%
\begin{equation}
\times \sqrt{\left( 2S_{0}+4m_{0}q\right) ^{2}-\left[ q\alpha (2n+1)+\sqrt{%
q^{2}\alpha ^{2}+4S_{0}^{2}}\right] ^{2}}.
\end{equation}%
In the above equation, all bound states appear in pairs, with energies $\pm
E_{n}.$ Since the Klein-Gordon equation is independent of the sign of $E_{n}$
for scalar potentials, the wavefunctions become the same for both energy
values. We notice that Eq. (82) is identical to Eq. (24) of Ref. [47]
obtained by N-U method and to Eq. (20) of Ref. [48] obtained by
supersymmetric method. If the range parameter $\alpha $ is chosen to be $%
\alpha =1/\lambda _{c},$ where $\lambda _{c}=\hbar /m_{0}c=1/m_{0}$ denotes
the Compton wavelength of the Klein-Gordon particle. It can be seen easily
that while $S_{0}\rightarrow 0$ in ground state $(n=0),$ all energy
eigenvalues tend to the value $E_{0}\approx 0.866$ $m_{0}.$

(ii) For $D=3,$ the mixed scalar and vector Hulth\'{e}n potentials $%
(q=1,l\neq 0)$, the energy eigenvalues are%
\begin{equation}
E_{nl}=\frac{\eta _{nl}V_{0}}{2}\pm \left[ \alpha (2n+1)+\sqrt{\alpha ^{2}+4B%
}\right] \sqrt{\frac{m_{0}^{2}}{4V_{0}^{2}+\left[ \alpha (2n+1)+\sqrt{\alpha
^{2}+4B}\right] ^{2}}-\left( \frac{\eta _{nl}}{4}\right) ^{2}},
\end{equation}%
where%
\begin{equation}
\eta _{nl}=\frac{4(V_{0}^{2}-S_{0}^{2})+\left[ \alpha (2n+1)+\sqrt{\alpha
^{2}+4B}\right] ^{2}-8m_{0}S_{0}}{4V_{0}^{2}+\left[ \alpha (2n+1)+\sqrt{%
\alpha ^{2}+4B}\right] ^{2}},\text{ }B=S_{0}^{2}-V_{0}^{2}+\alpha ^{2}l(l+1).
\end{equation}%
In order that at least one level might exist, it is necessary that the
inequality%
\begin{equation}
16m_{0}^{2}\geq \eta _{nl}^{2}\left\{ 4V_{0}^{2}+\left[ \alpha (2n+1)+\sqrt{%
\alpha ^{2}+4B}\right] ^{2}\right\} ,\text{ }\alpha ^{2}+4B\geq 0,
\end{equation}%
is fulfilled. In the case of pure vector ($S_{0}=0,V_{0}\neq 0),$the energy
eigenvalues are%
\begin{equation*}
E_{nl}=\frac{V_{0}}{2}\pm \left[ \alpha (2n+1)+\sqrt{\alpha
^{2}(2l+1)^{2}-4V_{0}^{2}}\right]
\end{equation*}%
\begin{equation}
\times \sqrt{\frac{m_{0}^{2}}{4V_{0}^{2}+\left[ \alpha (2n+1)+\sqrt{\alpha
^{2}(2l+1)^{2}-4V_{0}^{2}}\right] ^{2}}-\frac{1}{16}},
\end{equation}%
with the following constraint over the potential parameters:%
\begin{equation*}
16m_{0}^{2}\geq 4V_{0}^{2}+\left[ \alpha (2n+1)+\sqrt{\alpha
^{2}(2l+1)^{2}-4V_{0}^{2}}\right] ^{2},
\end{equation*}%
\begin{equation}
(2l+1)\alpha \geq 2V_{0}.
\end{equation}%
A preliminary analysis about the possibility of obtaining real bound states
for the Klein-Gordon particle of mass unity is done for various dimensions
as follows: when $D=3,$ we have only two lower-lying real bound states $%
(n=0,1)$ for the angular quantum number $l=0,$ one lower-lying state $(n=0)$
for $l=1$ and no states for $l\geq 2.$ When $D=4,$ we have two lower-lying
states for $l=0$ and no states for $l\geq 1.$ Further, when $D=5,$ we have
one lower-lying state for $l=0$ and no states for $l\geq 1$. Finally, when $%
D\geq 6,$ there are no real bound states for $l\geq 0,$ however, only
scattering states will be possible.

(iii) When $D=3$ and $l=0,$ the centrifugal term $\frac{(D+2l-1)(D+2l-3)}{%
4r^{2}}=0,$ and the approximation term $\frac{(D+2l-1)(D+2l-3)\alpha
^{2}e^{-\alpha r}}{4\left( 1-qe^{-\alpha r}\right) ^{2}}=0,$ too. Thus,
letting $l=0$ and $D=3$ in Eq. (10), it reduces to the exact spectrum
formula and normalized radial eigenfunctions of the Klein-Gordon equation
for vector and scalar general Hulth\'{e}n-type potentials:%
\begin{equation*}
\sqrt{m_{0}^{2}-E_{n}^{2}}=\frac{%
2qr_{0}(m_{0}S_{0}+E_{nl}V_{0})+r_{0}(S_{0}^{2}-V_{0}^{2})}{2q^{2}(n+\delta )%
}-\frac{n+\delta }{2r_{0}},
\end{equation*}

\begin{equation}
\delta =\frac{1}{2}\left[ 1+\frac{1}{q}\sqrt{%
q^{2}+4r_{0}^{2}(S_{0}^{2}-V_{0}^{2})}\right] ,\text{ (}n=0,1,2,3,\cdots )
\end{equation}%
which gives%
\begin{equation*}
E_{n}=\frac{\eta _{n}V_{0}}{2q}+\left( q\alpha (2n+1)+\sqrt{q^{2}\alpha
^{2}+4(S_{0}^{2}-V_{0}^{2})}\right)
\end{equation*}%
\begin{equation*}
\times \sqrt{\frac{m_{0}^{2}}{4V_{0}^{2}+\left( q\alpha (2n+1)+\sqrt{%
q^{2}\alpha ^{2}+4(S_{0}^{2}-V_{0}^{2})}\right) ^{2}}-\left( \frac{\eta _{n}%
}{4q}\right) ^{2}},\text{ (}n=0,1,2,3,\cdots )
\end{equation*}%
\begin{equation*}
16q^{2}m_{0}^{2}\geq \eta _{n}^{2}\left[ 4V_{0}^{2}+\left( q\alpha (2n+1)+%
\sqrt{q^{2}\alpha ^{2}+4(S_{0}^{2}-V_{0}^{2})}\right) ^{2}\right] ,\text{ }%
q^{2}\alpha ^{2}+4(S_{0}^{2}-V_{0}^{2})\geq 0,
\end{equation*}%
\begin{equation}
\eta _{n}=\frac{4(V_{0}^{2}-S_{0}^{2})+\left( q\alpha (2n+1)+\sqrt{%
q^{2}\alpha ^{2}+4(S_{0}^{2}-V_{0}^{2})}\right) ^{2}-8qm_{0}S_{0}}{%
4V_{0}^{2}+\left( q\alpha (2n+1)+\sqrt{q^{2}\alpha
^{2}+4(S_{0}^{2}-V_{0}^{2})}\right) ^{2}}.
\end{equation}%
(iv) In the case $S_{0}=V_{0}$ Hulth\'{e}n potential, Eqs. (67) and (68) can
be reduced to the relativistic energy equation:%
\begin{equation}
\sqrt{m_{0}^{2}-E_{R}^{2}}=\frac{r_{0}V_{0}(m_{0}+E_{R})}{(n+\delta )}-\frac{%
n+\delta }{2r_{0}},\text{ }\delta =\frac{(D+2l-1)}{2},\text{ }%
(n,l=0,1,2,\cdots ).
\end{equation}%
which is Eq. (22) of Ref. [19].

(v) We discuss non-relativistic limit of the energy equation (90). When $%
V_{0}=S_{0},$ Eq. (10) reduces to a Schr\"{o}dinger-like equation for the
potential $2V(r).$ In other words, the non-relativistic limit is the Schr%
\"{o}dinger equation for the potential $-2V_{0}e^{-r/r_{0}}/\left[
1-e^{-r/r_{0}}\right] .$ Hence, using the transformation $%
m_{0}+E_{R}\rightarrow 2m_{0}$ and $m_{0}-E_{R}\rightarrow -E_{NR}$ [22], we
obtain the non-relativistic energy equation:%
\begin{equation}
E_{NR}=-\frac{1}{8m_{0}\alpha ^{2}}\left[ \frac{4m_{0}V_{0}-\alpha
^{2}(n+\delta )^{2}}{(n+\delta )}\right] ^{2},\text{ }\alpha =r_{0}^{-1}
\end{equation}%
which is Eq. (23) of Ref. [19] with $\delta $ is given in Eq. (90). It is
worthwhile to remark that Eq. (91) is identical to Eq. (59) of Ref. [30]
when the potential is $2V(r)$ and $\alpha $ becomes pure imaginary, i.e., $%
\alpha \rightarrow i\alpha .$

Thus, in the weak coupling condition, $\left[ (n+\delta )/m_{0}r_{0}\right]
^{2}\ll 1,$ $\left[ V_{0}r_{0}/(n+\delta )\right] ^{2}\ll 1,$ expanding the
energy equation (90), retaining only the term containing the power of $%
(1/m_{0}r_{0})^{2}$ and $(r_{0}V_{0})^{4},$ we have the relativistic energy%
\begin{equation}
E_{R}\approx E_{NR}+m_{0}+4m_{0}\left( \frac{V_{0}r_{0}}{q(n+\delta )}%
\right) ^{4},
\end{equation}%
which is Eq. (24) of Ref. [19] with $\delta $ is given in Eq. (90). The
first term is the non-relativistic energy and third term is the relativistic
approximation to energy.

(vi) For a more specific case where $q=-1,$ the usual Hulth\'{e}n potential
is reduced to the shifted usual Woods-Saxon (WS) potential%
\begin{equation}
V(r)=-V_{0}+\frac{V_{0}}{1+e^{-r/r_{0}}},\text{ }S(r)=-S_{0}+\frac{S_{0}}{%
1+e^{-r/r_{0}}},\text{ }r_{0}=\alpha ^{-1},
\end{equation}%
and hence the energy eigenvalues, Eq. (69), for the general WS-type
potentials, i.e., $q\rightarrow -q$ and then $q=1,$ become%
\begin{equation*}
E_{nl}^{(D)}=-\frac{\xi _{nl}V_{0}}{2}\pm \left[ \sqrt{2\alpha
^{2}+4(S_{0}^{2}-V_{0}^{2})-\alpha ^{2}(D+2l-2)^{2}}-\alpha (2n+1)\right]
\end{equation*}%
\begin{equation*}
\text{ }\times \sqrt{\frac{m_{0}^{2}}{4V_{0}^{2}+\left[ \sqrt{2\alpha
^{2}+4(S_{0}^{2}-V_{0}^{2})-\alpha ^{2}(D+2l-2)^{2}}-\alpha (2n+1)\right]
^{2}}-\left( \frac{\xi _{nl}}{4}\right) ^{2}},
\end{equation*}%
\begin{equation}
16m_{0}^{2}\geq \xi _{nl}^{2}\left\{ 4V_{0}^{2}+\left[ \sqrt{2\alpha
^{2}+4(S_{0}^{2}-V_{0}^{2})-\alpha ^{2}(D+2l-2)^{2}}-\alpha (2n+1)\right]
^{2}\right\} ,
\end{equation}%
where
\begin{equation}
\xi _{nl}=\frac{4(V_{0}^{2}-S_{0}^{2})+\left[ \sqrt{2\alpha
^{2}+4(S_{0}^{2}-V_{0}^{2})-\alpha ^{2}(D+2l-2)^{2}}-\alpha (2n+1)\right]
^{2}+8m_{0}S_{0}}{4V_{0}^{2}+\left[ \sqrt{2\alpha
^{2}+4(S_{0}^{2}-V_{0}^{2})-\alpha ^{2}(D+2l-2)^{2}}-\alpha (2n+1)\right]
^{2}},
\end{equation}%
and the inequality
\begin{equation}
2\alpha ^{2}+4S_{0}^{2}\geq 4V_{0}^{2}+\alpha ^{2}(D+2l-2)^{2},
\end{equation}%
must be fulfilled. In the pure vector potential ($S_{0}=0,V_{0}\neq 0$), the
energy eigenvalues are%
\begin{equation*}
E_{nl}^{(D)}=-\frac{V_{0}}{2}\pm \left[ \sqrt{2\alpha ^{2}-\alpha
^{2}(D+2l-2)^{2}-4V_{0}^{2}}-\alpha (2n+1)\right]
\end{equation*}%
\begin{equation*}
\times \sqrt{\frac{m_{0}^{2}}{4V_{0}^{2}+\left[ \sqrt{2\alpha ^{2}-\alpha
^{2}(D+2l-2)^{2}-4V_{0}^{2}}-\alpha (2n+1)\right] ^{2}}-\frac{1}{16}},
\end{equation*}%
\begin{equation}
16m_{0}^{2}\geq 4V_{0}^{2}+\left[ \sqrt{2\alpha ^{2}-\alpha
^{2}(D+2l-2)^{2}-4V_{0}^{2}}-\alpha (2n+1)\right] ^{2},
\end{equation}%
where%
\begin{equation}
2\alpha ^{2}\geq 4V_{0}^{2}+\alpha ^{2}(D+2l-2)^{2}.
\end{equation}%
>From the above equation, for any given $\alpha ,$ the Klein-Gordon equation
with the usual shifted WS potential has negartive eigenvalues, i.e, $%
E_{nl}<0.$ The given restriction in (99) imposes the critical coupling value
and thus leads to the result%
\begin{equation}
n\leq \frac{1}{\alpha }\left( \sqrt{\frac{\alpha ^{2}}{4}\left[
2-(D+2l-2)^{2}\right] -V_{0}^{2}}-\sqrt{4m_{0}^{2}-V_{0}^{2}}\right) -\frac{1%
}{2},
\end{equation}%
i.e., there are only finitly many eigenvalues. Further, it is necessary that
the inequality:%
\begin{equation}
\sqrt{\alpha ^{2}\left[ 2-(D+2l-2)^{2}\right] -4V_{0}^{2}}\geq \alpha +2%
\sqrt{4m_{0}^{2}-V_{0}^{2}},
\end{equation}%
must be fulfilled.

Now, let us find the wave function $y_{n}(s),$ which is the polynomial
solution of hypergeometric-type equation, we multiply Eq. (19) by the weight
function $\rho (s)$ so that it can be rewritten in self-adjoint form [36]%
\begin{equation}
\left[ \omega (s)y_{n}^{\prime }(s)\right] ^{\prime }+\lambda \rho
(s)y_{n}(s)=0.
\end{equation}%
The weight function $\rho (s)$ which satisfies Eqs. (24) and (102) has the
form%
\begin{equation}
\rho (s)=s^{2\varepsilon _{nl}}(1-s)^{\sqrt{1+4(\beta _{2}+\gamma )}},
\end{equation}%
and consequently from the Rodrigues relation (23), we obtain%
\begin{equation*}
y_{nl}(s)=B_{nl}s^{-2\varepsilon _{nl}}(1-s)^{-\sqrt{1+4(\beta _{2}+\gamma )}%
}\frac{d^{n}}{ds^{n}}\left[ s^{n+2\varepsilon _{nl}}(1-s)^{n+\sqrt{1+4(\beta
_{2}+\gamma )}}\right]
\end{equation*}%
\begin{equation}
=B_{nl}P_{n}^{(2\varepsilon _{nl},\sqrt{1+4(\beta _{2}+\gamma )})}(1-2s).
\end{equation}%
On the other hand, inserting the values of $\sigma (s),\pi (s)$ and $\tau
(s) $ given in Eqs. (59), (63) and (64) into Eq. (20), one can find the
other part of the wave function as%
\begin{equation}
\phi (s)=s^{\varepsilon _{nl}}(1-s)^{\frac{1}{2}\left[ 1+\sqrt{1+4(\beta
_{2}+\gamma )}\right] }.
\end{equation}%
Hence, the wave function in Eq. (18) becomes%
\begin{equation*}
g(s)=C_{nl}s^{\varepsilon _{nl}}(1-s)^{\frac{1}{2}\left[ 1+\sqrt{1+4(\beta
_{2}+\gamma )}\right] }P_{n}^{(2\varepsilon _{nl},\sqrt{1+4(\beta
_{2}+\gamma )})}(1-2s)
\end{equation*}%
\begin{equation}
=C_{nl}s^{\varepsilon _{nl}^{(D)}}(1-s)^{\delta }P_{n}^{(2\varepsilon
_{nl}^{(D)},2\delta -1)}(1-2s),\text{ }s\in \lbrack q,0).
\end{equation}%
Finally, the radial wave functions of the Klein-Gordon equation are obtained
as%
\begin{equation}
R_{l}(r)=N_{nl}r^{-(D-1)/2}e^{-\sqrt{m_{0}^{2}-E_{nl}^{2}}r\text{ }%
}(1-qe^{-r/r_{0}})^{(q+a)/(2q)}P_{n}^{(2r_{0}\sqrt{m_{0}^{2}-E_{nl}^{2}}%
,a/q)}(1-2qe^{-r/r_{0}}),
\end{equation}%
where $N_{nl}$ is the radial normalization factor. Thus, the radial wave
functions for the $s$-wave Klein-Gordon equation with pure vector Hulth\'{e}%
n potential in $1D$ reduces to%
\begin{equation}
R_{n}(r)=C_{nl}e^{-\sqrt{m_{0}^{2}-E_{nl}^{2}}r\text{ }%
}(1-qe^{-r/r_{0}})^{(q+a)/(2q)}P_{n}^{(2r_{0}\sqrt{m_{0}^{2}-E_{nl}^{2}}%
,a/q)}(1-2qe^{-r/r_{0}}),
\end{equation}%
where $a=\sqrt{q^{2}-4V_{0}^{2}/\alpha ^{2}}$ which is identical to Eq. (35)
of Ref. [30]. Finally, from Eqs. (5) and (6), the total wave functions for
the usual Hulth\'{e}n potential is%
\begin{equation*}
\psi _{l_{1}\cdots l_{D-2}}^{(l_{D-1}=l)}(\mathbf{x})=N_{nl}r^{-(D-1)/2}e^{-%
\sqrt{m_{0}^{2}-E_{nl}^{2}}}(1-e^{-r/r_{0}})^{\delta }P_{n}^{(2r_{0}\sqrt{%
m_{0}^{2}-E_{nl}^{2}},2\delta -1)}(1-2e^{-r/r_{0}})
\end{equation*}%
\begin{equation*}
\frac{1}{\sqrt{2\pi }}\exp (\pm il_{1}\theta _{1})\prod\limits_{j=2}^{D-2}%
\sqrt{\frac{\left( 2l_{j}+j-1\right) n_{j}!}{2\Gamma \left(
l_{j}+l_{j-1}+j-2\right) }}\left( \sin \theta _{j}\right)
^{^{l_{j}-n_{j}}}P_{n_{j}}^{(l_{j}-n_{j}+(j-2)/2,l_{j}-n_{j}+(j-2)/2)}(\cos
\theta _{j})
\end{equation*}%
\begin{equation}
\sqrt{\frac{\left( 2n_{D-1}+2m^{\prime }+1\right) n_{D-1}!}{2\Gamma \left(
n_{D-1}+2m^{\prime }\right) }}\left( \sin \theta _{D-1}\right)
^{l_{D-2}}P_{n_{D-1}}^{(m^{\prime },m^{\prime })}(\cos \theta _{D-1}),
\end{equation}%
where
\begin{equation}
\delta =\frac{1}{2}\left( 1+\sqrt{%
(D+2l-2)^{2}+4r_{0}^{2}(S_{0}^{2}-V_{0}^{2})}\right) ,\text{ }%
(l=0,1,2,\cdots ).
\end{equation}

\section{Conclusions}

We have analytically found an approximate bound state energy\ eigenvalues
and their corresponding wave functions of the $D$-dimensional Klein-Gordon
equation for the spin-zero particle with the scalar and vector general Hulth%
\'{e}n-type potentials using the N-U method. The analytic energy equation
and the wave functions expressed in terms of Jacobi polynomials and can be
reduced to their well-known $1D$ and $3D$ solutions. The relativistic
energies $E_{nl}$ in $D$-dimensions defined explicitly in Eq. (68) is for
the particle and anti-particle energies for a given dimension $D,$ quantum
numbers $n$ and $l$ and also coupling constants satisfying the given
particular constraints. For pure attractive scalar potential, all bound
states appear in pairs, with energies $\pm E_{nl}.$ Since the Klein-Gordon
equation is independent of the sign of $E_{nl}$ for scalar potential, the
wave functions become the same for both energy values. When $l=0,$ the
results in this work reduce to exact solution of bound states of $s$-wave
Klein-Gordon equation with vector and scalar general Hulth\'{e}n-type
potentials. A preliminary analysis about the possibility of obtaining finite
number states for the Klein-Gordon particle of mass unity shows that the
bound state energies are positive for the general Hulth\'{e}n-type
potentials and negative for general Woods-Saxon-type potentials.

In Table 1, we have obtained numerical results for the binding energies of
ground state for mass unity $1D$ Klein Gordon equation with pure vector and
pure scalar cases. \ In pure vector potential, for fixed coupling constants $%
V_{0}=0.25$ $m_{0}$ and various values of $\alpha =0.5,1.0$ and $2.0,$ most
of the binding energies $E_{0}$ are decreasing with increasing deformation
constant $q.$ Also the values obtained with pure scalar potential, $%
S_{0}=0.25$ $m_{0},$ are also given in Table 1 for comparison. Furthermore,
in Table 2, we give the binding energies for the $n=1,$ $\alpha =0.5,1.0$
and $n=2,$ $\alpha =0.5.$ Obviously, we have no bound states for $n=1$ when $%
\alpha =2.0$ and for $n=2$ when $\alpha =1.0$ and $2.0.$

\acknowledgments Work partially supported by the Scientific and
Technological Research Council of Turkey (T\"{U}B\.{I}TAK).

\newpage

{\normalsize 
}

\newpage \baselineskip= 2\baselineskip
\bigskip \newpage

\begin{table}[tbp]
\caption{Relativistic ground state binding energies, $E_{0},$ for a particle
of mass, $m_{0}=1$ given by Eqs. (79) and (82) for pure vector and pure
scalar potentials, respectively, as a function of $q$ for various values of $%
\protect\alpha .$}%
\begin{tabular}{lllllll}
& Pure vector & ($V_{0}=0.25)$\tablenotemark[1]\tablenotetext[1]{For real
bound states $q^{2}$$\alpha^{2}$$\geq$$0.25$.} &  & Pure Scalar & ($%
S_{0}=0.25)$ &  \\
$q$ & $\alpha =0.5$ & $\alpha =1.0$ & $\alpha =2.0$ & $\alpha =0.5$ & $%
\alpha =1.0$ & $\alpha =2.0$ \\
\tableline$0.1$ & $-$ & $-$ & $-$ & $0.755639$ & $0.947484$ & $0.946410$ \\
$0.5$ & $-$ & $0.911438$ & $0.500000$ & $0.929812$ & $0.996314$ & $0.645297$
\\
$1.0$ & $0.820971$ & $0.970804$ & $0.250000$ & $0.986425$ & $0.964541$ & $%
0.480683$ \\
$1.5$ & $0.991673$ & $0.940945$ & $0.166683$ & $0.998606$ & $0.941246$ & $%
0.398341$ \\
$2.0$ & $0.999840$ & $0.923893$ & $0.125000$ & $0.999903$ & $0.926256$ & $%
0.347332$ \\
$2.5$ & $0.999140$ & $0.913089$ & $0.100000$ & $0.998350$ & $0.916087$ & $%
0.311871$ \\
$5.0$ & $0.988667$ & $0.890301$ & $0.050000$ & $0.988537$ & $0.892966$ & $%
0.222136$ \\
$7.5$ & $0.982893$ & $0.882371$ & $0.033334$ & $0.982996$ & $0.884414$ & $%
0.181787$ \\
$10$ & $0.979613$ & $0.878345$ & $0.025000$ & $0.979771$ & $0.879977$ & $%
0.157607$%
\end{tabular}%
\end{table}

\bigskip

\begin{table}[tbp]
\caption{Relativistic binding energies of the excited states, $E_{1}$ and $%
E_{2},$ for a particle of mass, $m_{0}=1$ given by Eqs. (79) and (82) for
pure vector and pure scalar potentials, respectively, as a function of $q$
for various values of $\protect\alpha .$}%
\begin{tabular}{lllllll}
& $n=1$\tablenotemark[1]\tablenotetext[1]{Pure vector case with coupling
constant $V_{0}=0.25$.} &  & $n=2$ & $n=1$\tablenotemark[2]%
\tablenotetext[2]{Pure scalar case with coupling constant $S_{0}=0.25$.} &
& $n=2$ \\
$q$ & $\alpha =0.5$ & $\alpha =1.0$ & $\alpha =0.5$ & $\alpha =0.5$ & $%
\alpha =1.0$ & $\alpha =0.5$ \\
\tableline$0.1$ & $-$ & $-$ & - & $0.995674$ & $0.771938$ & $0.922413$ \\
$0.5$ & $-$ & $0.830948$ & - & $0.984202$ & $0.571823$ & $0.829156$ \\
$1.0$ & $0.996421$ & $0.347292$ & $0.880588$ & $0.954903$ & $0.450227$ & $%
0.779514$ \\
$1.5$ & $0.949420$ & $0.229259$ & $0.769589$ & $0.935491$ & $0.381304$ & $%
0.752337$ \\
$2.0$ & $0.928534$ & $0.171407$ & $0.737131$ & $0.922604$ & $0.336188$ & $%
0.735135$ \\
$2.5$ & $0.916027$ & $0.136934$ & $0.719697$ & $0.913600$ & $0.303882$ & $%
0.723317$ \\
$5.0$ & $0.891025$ & $0.068342$ & $0.688449$ & $0.892282$ & $0.219316$ & $%
0.695608$ \\
$7.5$ & $0.882692$ & $0.045546$ & $0.678994$ & $0.884103$ & $0.180255$ & $%
0.684996$ \\
$10$ & $0.878525$ & $0.034156$ & $0.674438$ & $0.879801$ & $0.156613$ & $%
0.679406$%
\end{tabular}%
\end{table}

\end{document}